\documentclass[prb,twocolumn,showpacs,epsfig,epsf]{revtex4}
\usepackage{amsmath,amsthm,amsfonts,amscd,mathrsfs,pifont,bm}
\usepackage{eucal,graphicx}
\begin{document}

\title  {Note on the representation of the gap formation probability for real and quaternion Wishart matrices}

\author {Pedro A. Vidal Miranda}

\affiliation
       {
       }

\begin{abstract}
Wishart random matrices are often used to model multivariate systems in physics, finance, biology and wireless communication.
Extreme value statistics, such as those of the smallest eigenvalue, can be used to test the accuracy of the model. 
In this article we study the gap formation probability (cumulative distribution function of the smallest eigenvalue)
for real and quaternion $N\times \left(N+\nu\right)$ Wishart random matrices
in the large $N$ limit.
We derive compact expressions in terms of determinants of known functions.
As a consequence of these representations, the gap formation probabilities solve the Toda lattice equation, in the index
$\nu$ 
for $\nu$ even and for $\nu$ odd separately.
\end{abstract}


\maketitle
\newpage
{\bf Introduction.}---
Wishart random matrices with real, complex and quaternion entries are used to model the statistics of data and systems
 in a wide variety of disciplines.
They are used to model time series in financial data \cite{Muirhead,Anderson,Chatfield,Plerou},
human EEG data \cite{Seba}, 
the bipartite entanglement for a generic quantum system \cite{Vivo,Majumdar2} or the Hamiltonian of topological insulators.
In QCD they are referred to as Chiral Ensembles and they describe the Dirac spectrum
while in multivariate analysis they are used for principal components analysis 
of large data sets \cite{Majumdar}. In multivariate analysis the Wishart ensemble with correlation matrix equal to the identity
is called the null case and
knowledge about the null case allows one to perform tests
of the null hypothesis on data.

For most of these applications the matrices are real which makes the real Wishart model 
perhaps the most interesting and 
there has been a continued effort\cite{Tim2,Damgaard,Akemann,Vivo,Recher,Recher2,Chen} to characterize it.
One of the key quantities that has been studied is the distribution 
of the smallest eigenvalue.
In general, the smallest eigenvalue serves as an estimator for particular quantities
of interest under consideration. 
For example, in quantum mechanics, when considering the question of how entangled a generic bipartite system
is when in a random pure state,
the eigenvalues of the reduced density matrix of one of the systems determines the degree of entanglement\cite{Vivo,Majumdar2}.
If the smallest eigenvalue is zero the state space of the density matrix loses effectively a dimension
and is therefore the system is less entangled. If the smallest eigenvalue acquires its maximum value, 
all eigenvalues must be equal and therefore it is fully entangled.  
The smallest eigenvalue therefore gives information on the degree of entanglement.

Our focus will be on the compact representations 
of the distribution
of the smallest eigenvalue, $P^{(\beta)}_{\nu}(s)$
and  the gap formation probability, $Q^{(\beta)}_{\nu}(s)$, in the large $N$ limit,
for real ($\beta=1$) and quaternion ($\beta=4$) matrices. Since  $P^{(\beta)}_{\nu}(s)$
can be computed from $Q^{(\beta)}_{\nu}(s)$ by differentiation, we will only discuss the representation of the gap formation probability.
We will show that the gap formation probability $Q^{(\beta)}_{\nu}(s)$
can be represented in compact form as a determinant, for any integer $\nu$. 
The appearance of a determinant representation 
reflects strikingly different integrable properties then those known. 
As a result of this representation it will be clear that the nearest neighboring (in the index $\nu$)
gap formation probabilities, $Q^{(1)}_{\nu}(s)$ and $Q^{(1)}_{\nu\pm 2}(s)$, 
are linked together via the Toda lattice equation.

{\bf Known results.}---
Real Wishart random matrices have been extensively studied 
 with regards to the statistics of the smallest eigenvalue
and the gap formation probability and  there are a wide variety of results.
For $\nu$ odd,  $P^{(1)}_{\nu}(s)$
and  $Q^{(1)}_{\nu}(s)$ have a representation in terms of Hypergeometric function of matrix arguments  
\cite{Chen,Forrester2}
and a representation in terms of a Pfaffian was derived in \cite{Damgaard,Nagao}, where the dimension of the matrix in the Pfaffian
is $(\nu+1)\times (\nu+1)$.
It is not until recently
that a the case where $\nu$ is even was finally tackled \cite{Tim,Akemann}
and further Paffian forms were uncovered, where the dimension of the matrix in the Pfaffian 
is $\frac{\nu}{2}\times \frac{\nu}{2}$ ( $\left(\frac{\nu}{2}+1\right)\times \left(\frac{\nu}{2}+1\right)$) when $\frac{\nu}{2}$ is even(odd). 
Much is also known of the behavior of the gap probability,
$Q^{(\beta)}_{\nu}(s)$, in terms of
solutions to the Painlev\'e $V$ equation \cite{ForresterWitte,Forrester3}, i.e. 
 it was shown that, for arbitrary real $\nu$, $Q^{(\beta)}_{\nu}(s)$ solves a differential equation with some given boundary conditions.
Our analysis will be almost completely based on these results. 
In a nutshell we 
will put forward an ansatz for $Q^{(1)}_{\nu}(s)$ and $Q^{(4)}_{\nu}(s)$
and show they solve these differential equations with the same boundary conditions.

Although we analyze the case where the Wishart matrix average correlation function is the identity
it worthy to note that numerical evidence has shown \cite{Tim2,Akemann} that 
the distribution of the smallest eigenvalue is universal, meaning it remains unchanged 
even after introduction of non trivial correlations.

{\bf 
Representation of $Q^{(1)}_{\nu}(s)$ as an integral over the symplectic group for $\nu$ odd.}---
In the Wishart Ensembles the $N\times \left(N+\nu\right)$ random matrix $W$ has real, complex or quaternion ($\beta=1,2$ or $4$) 
entries which are Gaussian distributed.
\begin{align}
 P(WW^{\dagger})\sim e^{-\frac{1}{2}\text{Tr}\left[WW^{\dagger}C^{-1}\right]}
\end{align}
where $C$ is the average correlation matrix and taken to be $\mathbb{I}_{N}$ here.
The joint probability distribution function (j.p.d.f.) of the eigenvalues
of $WW^{\dagger}$
is known to be
\begin{align}
P(WW^{\dagger})=& \frac{1}{C_{N,\nu}}\left|\Delta_{N}(w_{k})\right|^{\beta}\prod_{j=1}^{N}e^{-\frac{\beta}{2}w_{j}}w_{j}^{\frac{\beta}{2}\left(\nu+1\right)-1}.
\label{eq:jpdf}
\end{align}
with the Vandermonde determinant given by
\begin{align}
 \Delta_{N}(w_{k})=\prod_{1\leq j<l\leq N}\left(w_{j}-w_{l}\right)
\end{align}
and with $\beta=1,4$ for the real and quaternion ensemble and $C_{N,\nu}$ the normalization constant.
This j.p.d.f. has also been studied for arbitrary real values of $\nu$ and
it is generally referred to as the Laguerre Ensemble. 
We will specify when $\nu$ is integer, odd or even , but in general view it as an arbitrary real number. 
We are interested in the gap formation probability, i.e. the probability that there are no eigenvalues in the interval $\left[0,s\right]$, denoted by $q^{(1)}_{N,\nu}(s)$,
\begin{align}
q^{(\beta)}_{N,\nu}(s)
=& \frac{1}{C_{N,\nu}}\prod_{j=1}^{N}\int_{s}^{\infty} dw_{j}\left|\Delta_{N}(w_{k})\right|e^{-\frac{\beta}{2}w_{j}}w_{j}^{\frac{\beta}{2}\left(\nu+1\right)-1}
\label{eq:gap}
\end{align}
and in particular the large $N$ limit of it
\begin{align}
 Q^{(1)}_{\nu}\left(s\right)=
\lim_{N\rightarrow \infty}  q^{(1)}_{N,\nu}\left(\frac{s}{4N}\right).
\end{align}
The distribution of the smallest eigenvalue, $P^{(1)}_{N,\nu}(s)$, is given through the derivative of the gap formation probability
\begin{align}
P^{(1)}_{N,\nu}(s)=-\frac{\partial}{\partial s}Q^{(1)}_{N,\nu}(s) \label{eq:pq}
\end{align}
and thus completely determined by $Q^{(1)}_{N,\nu}(s)$.
For brevity  and clarity we will use the following function of the gap formation probability
\begin{align}
 \mathcal{Q}_{\nu}^{(\beta)}\left(s\right)=Q^{(\beta)}_{\nu}\left(s^{2}\right)
\end{align}
since most equations acquire a simpler form  when written for $Q^{(\beta)}_{\nu}\left(s^{2}\right)$.
We note in passing that $ \mathcal{Q}_{\nu}^{(\beta)}\left(s\right)$ is the gap formation probability for the 
smallest eigenvalue of the Dirac spectrum. 

An often overlooked result, proven in  \cite{Forrester3} (Eq. (5.44)), is that for $\nu=2m+1$ odd
the gap formation probability is equal to an integral over the symplectic 
group.
\begin{align}
\mathcal{Q}^{(1)}_{2m+1}\left(s\right)
=& e^{-\frac{\beta s^2}{8}}
\int_{Sp(m)}dUe^{\frac{s}{2}\text{Tr}\left[U\right]}\label{eq:gapsymplectic}
\end{align}
where $Sp(m)$ is the symplectic group and the integration measure on the group is the Haar measure (normalized)
\footnote{Integrals over Symplectic and Orthogonal matrices are always normalized to $1$.}.
The proof consisted of showing 
that the same Painlev\'e $V$ equation
was satisfied by both the left 
hand side \cite{Forrester2} and the right hand side \cite{Forrester3}
. In addition it was shown, both sides have the same boundary conditions,
thereby proving Eq. (\ref{eq:gapsymplectic}).
Denoting by $e^{i\theta_{j}}$ the eigenvalues of $U$, in Eq. (\ref{eq:gapsymplectic}), and setting $\lambda_{j}=\cos\theta_{j}$,
we can write this integral as follows \cite{ForresterBook}
\begin{align}
\mathcal{Q}^{(1)}_{2m+1}\left(s\right)
=&\frac{e^{-\frac{\beta s^2}{8}}}{C_{m}} 
\prod_{j=1}^{m}\int_{-1}^{1}d\lambda_{j}
\left|\Delta_{m}(\lambda_{k})\right|^2 e^{s\lambda_{j}}\left(1-\lambda_{j}^2\right)^{\frac{1}{2}} 
\label{eq:Qoddjpdf}
\end{align}
with $C_{m}$ the normalization constant. 
$C_{m}$ will be used in general to denote the normalization constants 
but will not always be the same.
From this expression,
 and using the Andr\'eief-de
Bruijn integration theorem
, we clearly have a representation in terms of a determinant.
\begin{align}
\mathcal{Q}^{(1)}_{2m+1}\left(s\right)
&=\frac{e^{-\frac{\beta s^2}{8}}}{C_{m}} \det_{ 0\leq j,k\leq m-1}\left[
\left(\frac{\partial}{\partial s }\right)^{j+k}\left(\frac{\pi I_{1}(s)}{s}\right)\right] \label{eq:Qodd}
\end{align}
with
\begin{align}
 \int_{-1}^{1}d\lambda\left(1-\lambda^2\right)^{\frac{1}{2}}
e^{s\lambda}=\frac{\pi I_{1}(s)}{s}
\end{align}
We stress that this representation is valid for $\nu=2m+1$ odd.  The determinant in Eq. (\ref{eq:Qodd}) is a Hankel determinant, 
which is known to solve the Toda lattice equation.

{\bf Representation of $\mathcal{Q}^{(1)}_{\nu}(s)$ for arbitrary integer $\nu$.}---
Before discussing the $\nu$ even case we make the following observation about the $\nu$ odd case.
It is easily seen, by comparing the j.p.d.f. of the eigenvalues,
that the integral over the symplectic group, in Eq. (\ref{eq:Qoddjpdf}), is equal to one over the
orthogonal matrices with determinant equal to $-1$ and of even dimension equal to $2m+2$, $O^{-}_{2m+2}$.
It was already noted in \cite{ForresterBook}, that the j.p.d.f of the integral over these two ensembles are the same.
Concretely we have from \cite{ForresterBook}
\begin{align}
& \int_{O^{-}_{2m+2}}dOe^{\frac{s}{2}\text{Tr}\left[O\right]} \notag\\
 =&
 \frac{1}{C_{m}}\prod_{j=1}^{m}\int_{-1}^{1}d\lambda_{j}\left|\Delta_{m}(\lambda_{j})\right|^2
 e^{s\lambda_{j}}
 \left(1-\lambda_{j}^2\right)^{\frac{1}{2}}
.  \notag
\end{align}
The condition that the determinant of the orthogonal matrices be equal to $-1$ fixes one eigenvalues to $1$ and the other to $-1$.
We have therefore from Eq. (\ref{eq:Qoddjpdf}) and for $\nu=2m+1$ odd 
\begin{align}
\mathcal{Q}^{(1)}_{2m+1}\left(s\right)&=e^{-\frac{\beta s^2}{8}}  \int_{O^{-}_{2m+2}}dOe^{\frac{s}{2}\text{Tr}\left[O\right]}.
\end{align}
From this equation we put forward 
the ansatz that, for $\nu=2m$ even, the gap formation probability is equal to the integral over 
 orthogonal matrices with odd dimension and determinant equal to $-1$, i.e. as we decrease $\nu=2m+1$ by $1$ we also decrease the dimension of 
 the orthogonal matrix by $1$:
\begin{align}
\mathcal{Q}^{(1)}_{2m}\left(s\right)
&\stackrel{?}{=}
e^{-\frac{\beta s^2}{8}} \int_{O^{-}_{2m+1}}dOe^{\frac{s}{2}\text{Tr}\left[O\right]} . \label{eq:guess}
\end{align}
As was done in \cite{Forrester3}
one can prove Eq. (\ref{eq:guess}) by showing that the left and right hand side
solve the same Painlev\'e $V$ equation, with the same boundary conditions.
More specifically  it was shown\cite{ForresterWitte} that,
for \underline{arbitrary real $\nu$},  
the following function of the gap formation probability 
\begin{align}
F(s) =&s\frac{\partial}{\partial s}\log \mathcal{Q}^{(1)}_{\nu}\left(s\right)
\end{align}
is in fact related to the Painlev\'e $V$ solution through
\begin{align}
F(s) =&\sigma_{V}(s)-\frac{s^2}{4}+\frac{\nu-1}{2}s-\frac{\nu(\nu-1)}{4} \label{eq:F}
\end{align}
with $\sigma_{V}(s)$ solving the following Painlev\'e $V$ equation 
with $s\rightarrow 2t$
\begin{align}
& \left(t\sigma''\right)^2-\left(\sigma-s\sigma'+2\left(\sigma'\right)^2+\mu\sigma'\right)^{2}\notag\\
& +4
 \left(\mu_{0}+\sigma'\right)
 \left(\mu_{1}+\sigma'\right)
 \left(\mu_{2}+\sigma'\right)
 \left(\mu_{3}+\sigma'\right)=0 \label{eq:Pequation}
\end{align}
and with the following coefficients
\begin{align}
\begin{array}{ccc}
\mu=\nu-1 &\mu_{0} =0  & \mu_{1}=\frac{\nu}{2}\\
 \mu_{2}=\frac{\nu-1}{2}& \mu_{3}=-\frac{1}{2} & \\
\end{array}.\label{eq:Pconditions}
\end{align}
The boundary condition is  given by 
\begin{align}
\lim_{s\rightarrow 0}F(s)
=&-\frac{s}{2}J_{\nu}(s)-\frac{s^2}{4}\left(J^{2}_{\nu}(s)-J_{\nu-1}(s)J_{\nu+1}(s)\right)
\notag\\
=&
-\left(\frac{s}{2}\right)^{\nu+1}\frac{1}{\nu!}
\notag
\end{align}
Therefore to prove Eq. (\ref{eq:guess}) it suffices to show that the right-hand side also solves 
this equation and with the same boundary condition.
The integrals over $O_{2m+1}^{(-)}$ of Eq. (\ref{eq:guess}) 
was studied in \cite{vanMoerbekeAdler}. Using the theory of $\tau$-functions it was shown they
are related to the solutions of the Painlev\'e $V$ equation with particular boundary conditions.
Using these results and comparing them with the previous ones from \cite{ForresterWitte},
it is straightforward to prove
the ansatz of Eq. (\ref{eq:guess}).
Combining the results for the $\nu$ odd and even cases we have 
for $\nu$ integer the following result:
\begin{align}
\mathcal{Q}_{ \nu}^{(1)}\left(s\right)
&=e^{-\frac{\beta s^2}{8}} \int_{O^{-}_{\nu+1}}dOe^{\frac{s}{2}\text{Tr}\left[O\right]} . \label{eq:evenodd}
\end{align}
For the case where $\nu=2m$ is even we obtain from the right-hand side of Eq. (\ref{eq:evenodd}) 
the following integral representation \cite{vanMoerbekeAdler,ForresterBook} 
\begin{align}
\mathcal{Q}_{2m}^{(1)}\left(s\right)
 &=\frac{e^{-\frac{s}{2}}}{C_{m}}\prod_{j=1}^{m}\int_{-1}^{1}d\lambda_{j}
 \left|\Delta_{m}(\lambda_{j})\right|^2e^{s\lambda_{j}}\left(\frac{1+\lambda_{j}}{1-\lambda_{j}}\right)^{\frac{1}{2}}
\end{align}
where the eigenvalues of $O$ are denoted by $\left\{e^{i\theta_{j}},e^{-i\theta_{j}}\right\}_{j=1,\cdots m}$ and $-1$, 
 and as before we have $\lambda_{j}=\cos\theta_{j}$.
For $\nu=2m$ even we have then 
\begin{align}
\mathcal{Q}_{2m}^{(1)}\left(s\right)
&=\frac{e^{-\frac{\beta s^2}{8}} e^{-\frac{s}{2}}}{C_{m}}
\det_{0\leq j,k\leq m-1}\left[
\left(\frac{\partial}{\partial s}\right)^{j+k}g(s)
\right] \label{eq:Qeven}
\end{align}
with 
\begin{align}
 g(s)=&\int_{-1}^{1}d\lambda\left(\frac{1+\lambda}{1-\lambda}\right)^{\frac{1}{2}}
e^{s\lambda} \notag\\
=&\pi \left(I_{0}(s)+ I_{1}(s)\right).\notag
\end{align}
Once again the determinant in Eq. (\ref{eq:Qeven}) is a Hankel determinant which solves the Toda Lattice equation.
Thus a byproduct of the representations given in Eqs. (\ref{eq:Qodd}) and (\ref{eq:Qeven}) 
 is that the gap formation probability, $\mathcal{Q}_{\nu}\left(s\right)$, 
satisfy a type of Toda Lattice equation.
For the gap formation probability we have 
\begin{align}
 &
4\partial_{s}^2 \log \mathcal{Q}_{\nu}\left(s\right)
= 
\frac{\mathcal{Q}_{\nu-2}\left(s\right)
\mathcal{Q}_{\nu+2}\left(s\right)
}{\mathcal{Q}_{\nu}^2\left(s\right)}
-1 .\notag
\end{align}
Notice how the equation separates between the odd and even cases of $\nu$.

{\bf Quaternion Wishart matrices.}---
We can proceed in an similar way for quaternion Wishart matrices.
Let us recall the known results.
It was proved \cite{ForresterWitte} that for arbitrary real $\nu$, $\mathcal{Q}_{\nu}^{(4)}\left(s\right)$
splits into the sum two $\tau$-functions
\begin{align} 
\mathcal{Q}_{\nu}^{(4)}\left(\frac{s}{2}\right)
&=\frac{1}{2}\left(\tau_{V}^{+}(s)+\tau^{-}_{V}(s)\right)\label{eq:Qtau}
\end{align}
such that $F^{\pm}(s)$, given by
\begin{align}
F^{\pm}(s)=s\partial_{s}\log \tau^{\pm}_{V}(s), 
\end{align}
is related to a solution of the Painlev\'e $V$ equation, Eq. (\ref{eq:Pequation}), with coefficients (\ref{eq:Pconditions}),
through the formula (\ref{eq:F}), with the following change $\nu\rightarrow 2\nu$ in the coefficients of Eqs.(\ref{eq:Pconditions}) . 
The two solutions differ in their boundary conditions
\begin{align}
\lim_{s\rightarrow 0}F^{\pm}(s)
=&
\pm\left(\frac{s}{2}\right)^{2\nu+1}\frac{1}{(2\nu)!}
\label{eq:Fboundary}.
\end{align}
For 
$2\left(\nu+1\right)-1=2m$ a compact representation of $\mathcal{Q}_{\nu}^{(4)}\left(s\right)$ in terms of a determinant
was derived\cite{Forrester3}.
However the condition on $\nu$ implies $\nu$ is a half-integer and therefore 
this case does not include the quaternion Wishart matrices (we recall $\nu$ is the difference between the amount of columns and rows and thus an integer). 
For $\nu=\frac{2m+1}{2}$ a half integer it was shown \cite{Forrester3}
that
\begin{align} 
\mathcal{Q}_{\nu=\frac{2m+1}{2}}^{(4)}\left(\frac{s}{2}\right)
&=e^{-\frac{s^2}{8}}\int_{O_{2m+2}}dOe^{\frac{s}{2}\text{Tr}\left[O\right]}.\label{eq:quatortho} 
\end{align}
By decomposing the integral in
 two integrals, one over orthogonal matrices with positive determinant 
 and one over orthogonal matrices with negative determinant,  
\begin{align} 
\mathcal{Q}_{\frac{2m+1}{2}}^{(4)}\left(\frac{s}{2}\right)
&=\frac{e^{-\frac{s^2}{8}}}{2}\left(\int_{O^{+}_{2m+2}}dOe^{\frac{s}{2}\text{Tr}\left[O\right]}
+\int_{O^{-}_{2m+2}}dOe^{\frac{s}{2}\text{Tr}\left[O\right]}
\right)
\end{align}
it was shown that these corresponded to the $\tau$-functions
$\tau^{\pm}_{V}$ solving the Painlev\'e equation with the appropriate boundary conditions.
Following suit, we put forward the ansatz 
\begin{align} 
\mathcal{Q}_{\nu=m}^{(4)}\left(\frac{s}{2}\right)\stackrel{?}{=}e^{-\frac{s^2}{8}}\int_{O_{2m+1}}dOe^{\frac{s}{2}\text{Tr}\left[O\right]}. 
\end{align}
and split this integral into the integrals over two ensembles
\begin{align} 
\mathcal{Q}_{m}^{(4)}\left(\frac{s}{2}\right)
&=\frac{e^{-\frac{s^2}{8}}}{2}\left(\int_{O^{+}_{2m+1}}dOe^{\frac{s}{2}\text{Tr}\left[O\right]}
+\int_{O^{-}_{2m+1}}dOe^{\frac{s}{2}\text{Tr}\left[O\right]}
\right).\label{eq:guess2}
\end{align}
As done previously, we can use the results \cite{vanMoerbekeAdler} to show 
both of these integrals satisfy 
the Painlev\'e $V$ equation with the appropriate boundary conditions given by Eq. (\ref{eq:Fboundary}).
The sign of the orthogonal matrix ensemble over which is integrated
corresponds then to the upper index of the $\tau$-function defined in Eq.(\ref{eq:Qtau})
and determines the boundary conditions.
We note that the second integral in Eq. (\ref{eq:guess2}) 
corresponds to our previous ansatz Eq. (\ref{eq:guess}).
Finally this leads to 
\begin{align} 
\mathcal{Q}_{m}^{(4)}\left(\frac{s}{2}\right)
=&
\frac{e^{-\frac{s^2}{8}}}{2}\left(\mathcal{Q}^{(1)}_{2m}\left(s\right)+\mathcal{Q}^{(1)}_{2m}\left(-s\right)\right) .\label{eq:Qdet}
\end{align}

We have compared our results with
various known particular cases (see \cite{Vivo} and references therein)
and have found them to agree.

{\bf Conclusions.}---
We have derived, utilizing known results form the literature, compact closed representations 
as a determinant of known functions,
for the 
gap formation probability in the large $N$ limit, for 
real Wishart matrices of size $N\times N+\nu$ .
We have also shown that the gap formation probability solves a Toda lattice equation in the index $\nu$
for $\nu$ even and for $\nu$ odd separately.
In the quaternion case, for which no previous results exist,
we have shown that it can be written as the sum of two determinants
each of which satisfies the Toda Lattice with different initial conditions.
Although the recently derived representations \cite{Tim,Akemann} of the gap 
formation probability are quite compact, they do not show 
Toda Lattice relationship between different indexes.

It is interesting to note 
that for $\nu$ even
the gap formation probability can be written as 
the average of
half integer powers of a characteristic polynomial.
Although half integer powers of 
 characteristic polynomials appear in many physics applications \cite{Fyodorov}
not much is known about their integrability properties, if they even have any.
Thus our results provide an example of such properties, albeit a simple one.

The determinant representation is also interesting from   
the perspective of representations of Hypergeometric Functions 
of scalar Matrix Argument. 
To our knowledge there is only a Pfaffian representation\cite{Gupta}
available for this type of Hypergeometric Function of Matrix Argument.

\end{document}